\documentclass[aps,prl,reprint,superscriptaddress,preprintnumber,showpacs]{revtex4-1}
\usepackage{amsmath}
\usepackage{amssymb}
\usepackage{graphics}
\usepackage{epsfig}
\usepackage{CJK}
\usepackage{color}

\begin{document}

\begin{CJK*}{GBK}{}


\title{Anisotropic giant magnetoresistance in NbSb$_2$}
\author{Kefeng Wang}
\altaffiliation{Present address: CNAM, Department of Physics, University of Maryland, College Park, Maryland 20742, USA}
\affiliation{Condensed Matter Physics and Materials Science Department, Brookhaven National Laboratory, Upton New York 11973 USA}
\author{D. Graf}
\affiliation{National High Magnetic Field Laboratory, Florida State University, Tallahassee, Florida 32306-4005, USA}
\author{Lijun Li}
\author{C. Petrovic}
\affiliation{Condensed Matter Physics and Materials Science Department, Brookhaven National Laboratory, Upton New York 11973 USA}

\date{\today}

\begin{abstract}
We report large transverse magnetoreistance (the magnetoresistant ratio $\sim 1.3\times10^5\%$ in 2 K and 9 T field, and $4.3\times 10^6\%$ in 0.4 K and 32 T field, without saturation) and field-induced metal-semiconductor-like transition in NbSb$_2$. Magnetoresistance is significantly suppressed but the metal-semiconductor-like transition persists when the current is along the $ac$-plane. The sign reversal of the Hall resistivity and Seebeck coefficient in the field, plus the electronic structure reveal the coexistence of a small number of holes with very high mobility and a large number of electrons with low mobility. The large MR is attributed to the change of the Fermi surface induced by the magnetic field in addition to the high mobility metal.
\end{abstract}

\maketitle
\end{CJK*}

The magnetic field response of the transport properties of condensed matter, especially the magnetoresistance (MR) (such as the giant magnetoresistance in magnetic multilayers and colossal magnetoresistance in manganites)\cite{mr1,mr2}, gives information about the characteristics of the Fermi surface \cite{mr3} and provides promising candidates for magnetic memory or other spintronic devices \cite{mr44}. Correspondingly the exploratory search for the new materials exhibiting extraordinary response to the magnetic field (such as the large MR) keeps being a central topic of the condensed matter physics and material science. Unlike the magnetic mechanism of the MR in manganites and mangetic multilayers, the large positive magnetoresistance was discovered in some nomagnetic metals (such as Ag$_{2-\delta}$Te/Se \cite{agte1,agte2}, topological insulator Bi$_2$Te$_3$ \cite{qt1,qt2,qt3}, semimetal Bi \cite{bi}, graphite \cite{graphite1,graphite2,graphite3} andCd$_3$As$_2$ \cite{Cd3As2}, as well as (Sr/Ca)MnBi$_2$ \cite{SrMnBi1,SrMnBi2}). These large MR effects are in contrast to the normal MR in simple metals which is usually very small because semiclassical transport gives quadratic field-dependent MR in the low field range which would saturate in the high field \cite{mr3}. Some MR (Ag$_{2-\delta}$Te/Se and topological insulator Bi$_2$Te$_3$) were related to the Dirac fermions with linear energy dispersion in the quantum limit \cite{quantummr}, while some (such as Bi) were related to the Fermi surface (FS) compensation of the semimetals.

Besides the change in the value of resistivity, the magnetic field could also induce the change in the temperature-dependent behavior of the resistivity, and then induce the nometallic behavior. These transition or crossover usually is accompanied by the extremely large MR. A typical example is the magnetic field induced melting of the charge-ordered state and correspondingly colossal magnetoresistance in manganite \cite{mr2,mit}. Recently, the metal-semiconductor crossover induced by magnetic field was observed in several nomagnetic metals/oxides such as PtSn$_4$ \cite{ptsn4}, PdCoO$_2$ \cite{PdCoO2-1}, WTe$_2$ \cite{WTe2} and semimetal graphite \cite{graphite1,graphite2,graphite3} and Bi \cite{bi}. This was accompanied by the extremely large magnetoresistant effect where the magnetoresistant ratio $MR=(\rho(H)-\rho(0))/\rho(0)$ approaches $\sim 10^5\%$ at low temperature. These extraordinary metal-semiconductor crossover and extremely large MR effects induces the revival of the study on the magnetic field response of the transport behavior and the detailed mechanism remains unclear. Since PdCoO$_2$ and WTe$_2$ has layered structure and the large MR is related to the quasi-two-dimensional (quasi-2D)FS \cite{PdCoO2-1,WTe2}, it is important to explore new materials that crystalline in different crystal structure but host similar phenomena. New materials and mechanisms of large MR are of high interest.

Here we report the extremely large MR and the possible magnetic field induced semiconducting gap in NbSb$_2$ single crystal. Transverse $MR$ approaches $1.3\times10^5\%$ in 2 K and 9 T, and $4.3\times 10^6\%$ in 0.4 K and 32 T field without saturation, with electric current parallel to the $b$-axis. The large MR is significantly suppressed but the metal-semiconductor-like transition persists when the current is along the $ac$-plane. The first-principle electronic structure calculation and quantum oscillations reveal two types of carrier pockets. Their distinct density and mobility induces the sign reversal of the Hall resistivity. The large MR effects are attributed to the change of the Fermi pocket with high mobility induced by magnetic field, in addition to orbital MR expected for high mobility metals.

Single crystals of NbSb$_2$ used in this study were grown using a high-temperature self-flux method \cite{crystal,NbSb2}. Nb (99.99$\%$) and excess Sb (99.99$\%$) with ratio Nb:Sb=1:19 were sealed in a quartz tube, heated to 1100 $^{\circ}C$ and slowly cooled to 650 $^{\circ}C$ where the crystals were decanted. X-ray diffraction (XRD) data were taken with Cu K$_{\alpha}$ ($\lambda=0.15418$ nm) radiation of Rigaku Miniflex powder diffractometer at the room temperature. Electrical transport measurements up to 9 T were conducted on polished samples in Quantum Design PPMS-9 with conventional four-wire method. High field MR and SdH oscillation up to 35 T were measured at National High Magnetic Field Laboratory in Tallahassee. Thermal transport properties were measured in Quantum Design PPMS-9 from 2 K to 350 K using one-heater-two-thermometer method. The direction of heat and electric current transport was along the $ab$-plane of single grain crystals with magnetic field along the \textit{c}-axis and perpendicular to the heat/electrical current. The relative error in our measurement was $\frac{\Delta \kappa}{\kappa}\sim$5$\%$ and $\frac{\Delta S}{S}\sim$5$\%$ based on Ni standard measured under identical conditions. Fist principle electronic structure calculation were performed using experimental lattice parameters within the full-potential linearized augmented plane wave (LAPW) method \cite{wien2k1} implemented in WIEN2k package.\cite{wien2k2} The general gradient approximation (GGA) of Perdew \textit{et al}.,\cite{gga} was used for exchange-correlation potential. The LAPW sphere radius were set to 2.5 Bohr for all atoms. The converged basis corresponding to $R_{min}k_{max}=7$ with additional local orbital were used where $R_{min}$ is the minimum LAPW sphere radius and $k_{max}$ is the plane wave cutoff.

\begin{figure}[tbp]
\includegraphics[scale=0.35]{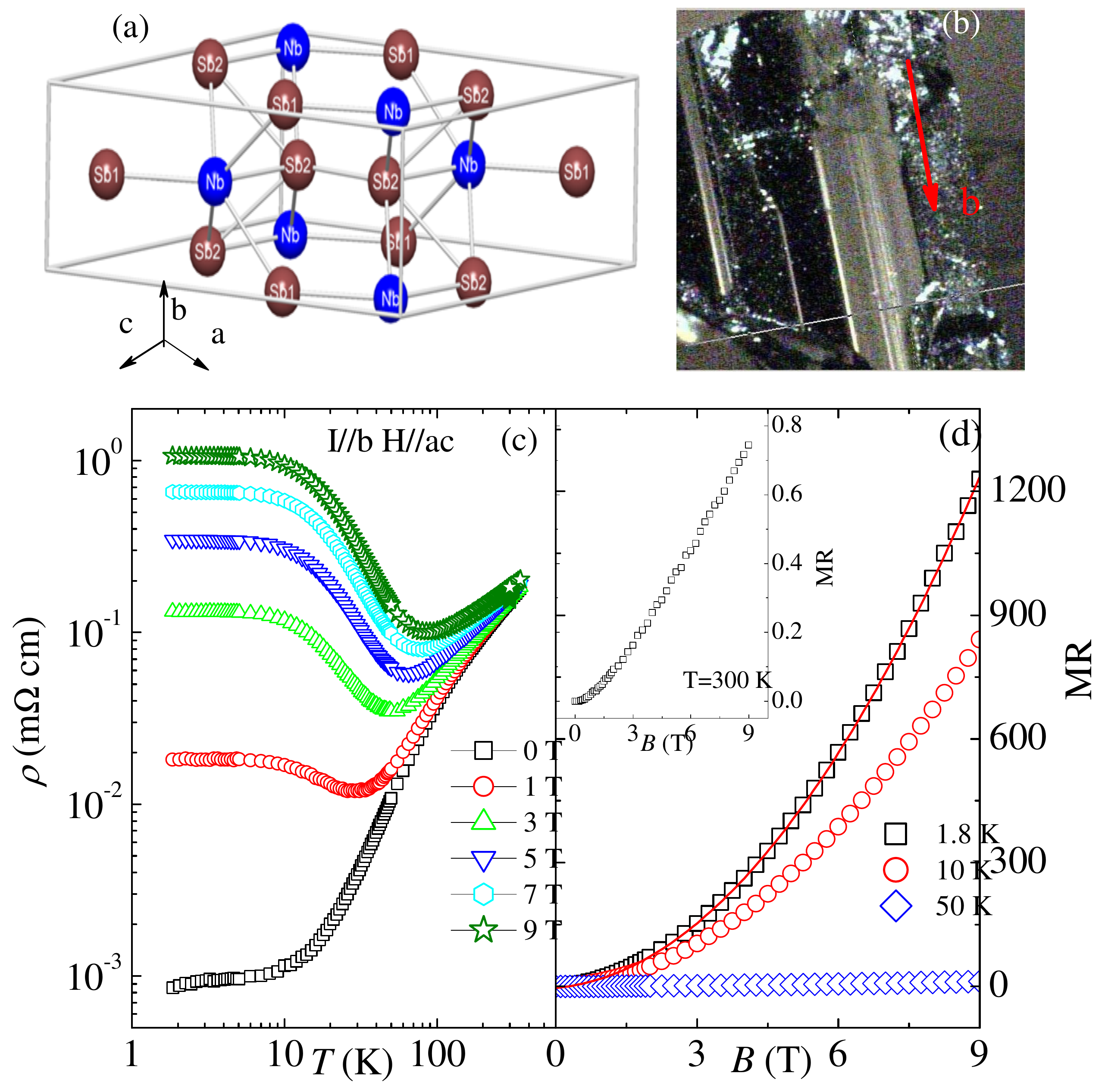}
\caption{(color online) \textbf{Crystal structure and giant magnetoresistance of NbSb$_2$.} (a) Crystal structure of NbSb$_2$. The angle $\beta$ is between the $a$ and $c$ axis. (b) A typical crystal of NbSb$_2$. The long axis of the crystal is $b$-axis. (c) The temperature dependence of the resistivity in magnetic fields with the current parallel to $b$-axis and the magnetic field perpendicular to the current (parallel to the $ac$-plane). (d) The magnetic field dependence of the magnetoresistant ratio defined as MR=(R(H)-R(0)/R(0), with same configuration as in (a). The red line is the fitting result using quadratic field dependence.}
\end{figure}

NbSb$_2$ crystallizes in a complex monoclinic structure (Fig. 1(a)) with C12/m1 space group with the refined lattice parameters are $a=10.233(1){\AA}, b=3.630(1){\AA}, c=8.3285(2){\AA}, \beta=120.04(2)^o$. 
The $b$-axis is perpendicular to the $ac$-plane (Fig. 1(b)). The image of a typical single crystal of NbSb$_2$ is shown in Fig. 1(c). 
The temperature dependent resistivity in different magnetic field of NbSb$_2$ with the current parallel to the $b$-axis and the magnetic field perpendicular to the current (parallel to the $ac$-plane) is shown in Fig. 1(c). The crystal shows metallic behavior and the residual resistivity ratio ($\rho_{300K}/\rho_{2K}$) in zero field is about 450 (with current along the $b$-axis) which suggests high sample quality.

\begin{figure}[tbp]
\includegraphics[scale=0.35] {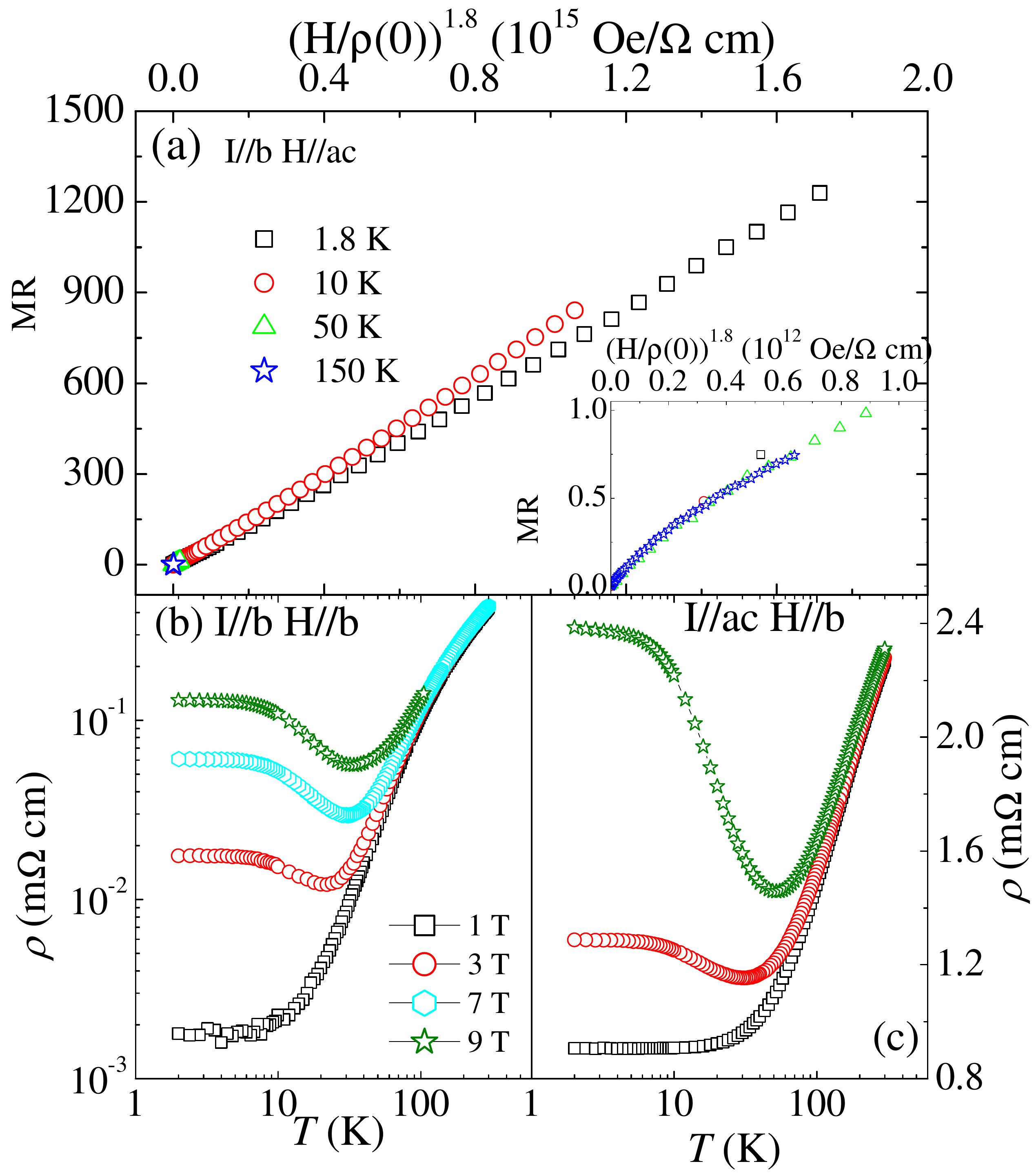}
\caption{(color online) The anisotropic giant magnetoresistance of NbSb$_2$. (a) A Kohler plot using $MR=F[H/\rho(0)]\simeq H^{1.8}$ for NbSb$_2$ with the current parallel to the $b$-axis and the magnetic field perpendicular to the current. Inset shows low field part enlarged for clarity. (b) The $\rho(T)$ with the current parallel to $b$-axis and the magnetic field parallel to the current (parallel to the $b$-axis. (c) The $\rho(T)$ with the current parallel to $ac$-plane and the magnetic field perpendicular to the current (parallel to the $b$-axis. }
\end{figure}

The external magnetic field significantly enhances the low-temperature resistivity and also changes its temperature-dependence. The $\rho(T)$ is metallic in 1 T [Fig. 1(c)] but the slope increases with temperature decrease from 300 K. At about 30 K the $\rho(T)$ shows a minimum and then increases with further decrease in temperature, which is similar to the metal-semiconductor crossover induced by external field observed in PdCoO$_2$ and WTe$_2$ \cite{PdCoO2-1,WTe2}. The resistivity nearly saturates below 10 K and its value increases from 0.009 $m\Omega$ cm in zero field to 0.02 $m\Omega$ cm in 1 T field. Higher field induces higher metal-semiconductor crossover temperature and larger MR. In 9 T field, the MR ratio approaches $\sim 1.3\times10^5\%$ in 2 K (as shown in Fig. 1(d)) and the metal-semiconductor crossover temperature is about 70 K. The MR does not saturate even in 35 T field. Instead it shows the quantum oscillations where $MR$ approaches $4.3\times 10^6\%$ in 0.4 K and 32 T field [Fig. 3(a) inset]. The magnetic field dependence of the MR (as shown in Fig. 1(d)) can be described very well by the parabolic behavior (the red line is the fitting result using $MR=aB^2$). With increasing temperature, $MR$ is suppressed significantly but the ratio is still around $80\%$ at 300 K [Fig. 1(d) inset]. Magnetotransport in semiclassical single-band metals scales as MR $=f(H\tau)=F(H/\rho_0)$ with the assumption of the single scattering time $\tau$, i.e. $1/\tau(T)\propto\rho_0(T)$ \cite{kohler}. MR of NbSb$_2$ in high field shows $MR\simeq H^{1.8}$ but deviates somewhat from Kohler scaling at low temperatures (Fig. 2(a)). The low field MR (inset of Fig. 2(a)) does not follow $H^{1.8}$ dependence but overall satisfies the scaling.

\begin{figure}[tbp]
\includegraphics[scale=0.35] {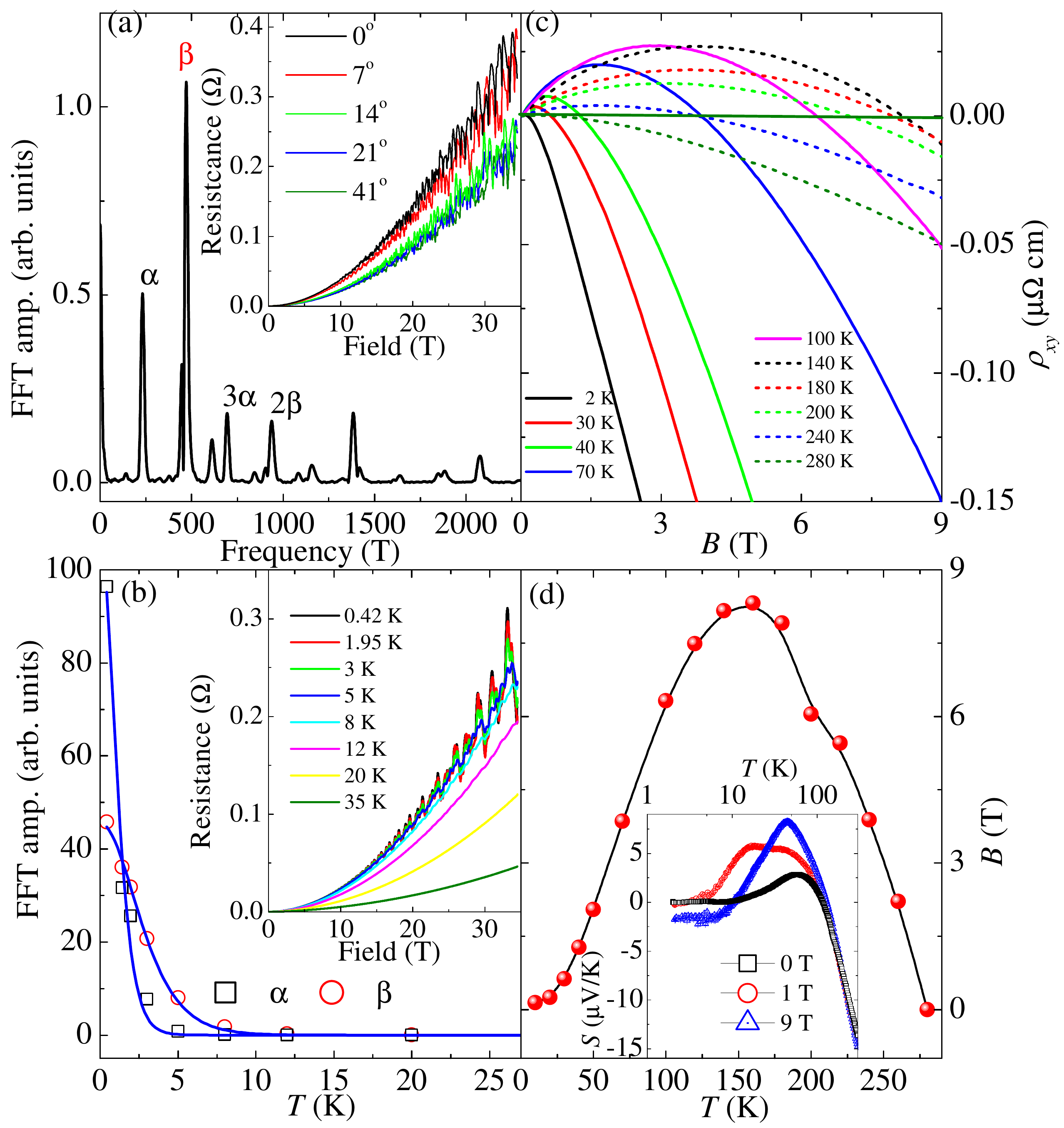}
\caption{(color online) (a) The FFT spectrum of the quantum oscillation of NbSb$_2$ crystal. The inset shows the typical SdH oscillation of NbSb$_2$ in 0.42 K with magnetic field up to 32 T. The current is parallel to $b$-axis and the magnetic field direction changes between perpendicular to the current (parallel to the $ac$-plane 0 degree) and perpendicular to $ac$-plane (90 degree). (b) The temperature dependence of the quantum oscillation amplitude for two oscillation frequencies. The red lines are the fit results which give the effective mass. The inset shows the magnetic field dependence of the oscillation at different temperatures. (c) The magnetic field dependence of the Hall resistivity $\rho_{xy}$ at different temperatures. (d) Sign change of Hall coefficient at different temperatures and magnetic fields. The inset shows the temperature dependence of the Seebeck coefficient $S$ with different magnetic field.} 
\end{figure}

The $\rho(T)$ and MR of NbSb$_2$ is significantly dependent on the electric current and magnetic field direction. When the electric current is still along the $b$-axis but the magnetic field is parallel to the current, the MR ratio is suppressed and is around $10^4\%$ in 9 T field and 2 K, as shown in Fig. 2(b). The anisotropic ratio ($\delta=\rho(B//b)/\rho(B//ac)$) is around 20 in 2 K and 9 T field, which is significantly larger than the value of both ordinary nomagnetic and magnetic metals (the ratio is around $1\%$ for alkali metals and $3\%$ for magnetic metals) \cite{mr3,mr4}. The resistivity along $ac$-plane (Fig. 2(c)) is three orders of magnitude higher than the resistivity along $b$-axis.  With the electric current along $ac$-plane and the field perpendicular to the current (parallel to $b$-axis) the MR ratio is suppressed and is only $300\%$ in the same temperature and field (Fig. 2(c)). However, the metal-semiconductor crossover is preserved and the crossover temperature is nearly the same in same magnetic field, in all configurations.

The highly anisotropic magnetotransport behavior usually implies anisotropic FS and is not expected in NbSb$_2$ crystal since the crystal structure is three-dimensional (Fig. 1(a) and Ref.\cite{NbSb2}). Fig. 3(a) and (b) shows the results of the Shubnikov-De Haas (SdH) oscillation for NbSb$_2$ crystals where the current is along $b$-axis but the field direction changes between parallel to the $ac$-plane (perpendicular to $b$-axis,$\theta=0^o$) and perpendicular to $ac$-plane ($\theta=90^o$). The MR of NbSb$_2$ shows clear oscillations above 8 T field (inset in Fig. 3(a)), and the oscillation is very clear in the whole angle range implying dominant 3D FS. The fast Fourier transformation (FFT) spectrum of oscillation (Fig. 3(a)) shows two major peaks at 227 T and 483 T oscillation frequency. The temperature dependence of the oscillation amplitude (Fig. 3(b)) can be fitted well by Lifshitz-Kosevitch formula \cite{osc1}, which gives the effective cyclotron resonant mass $m^*=0.68m_e$ and $1.69m_e$ ($m_e$ is the mass of bare electron) for 227 T and 483 T pockets respectively.

\begin{figure}[tbp]
\includegraphics[scale=0.32] {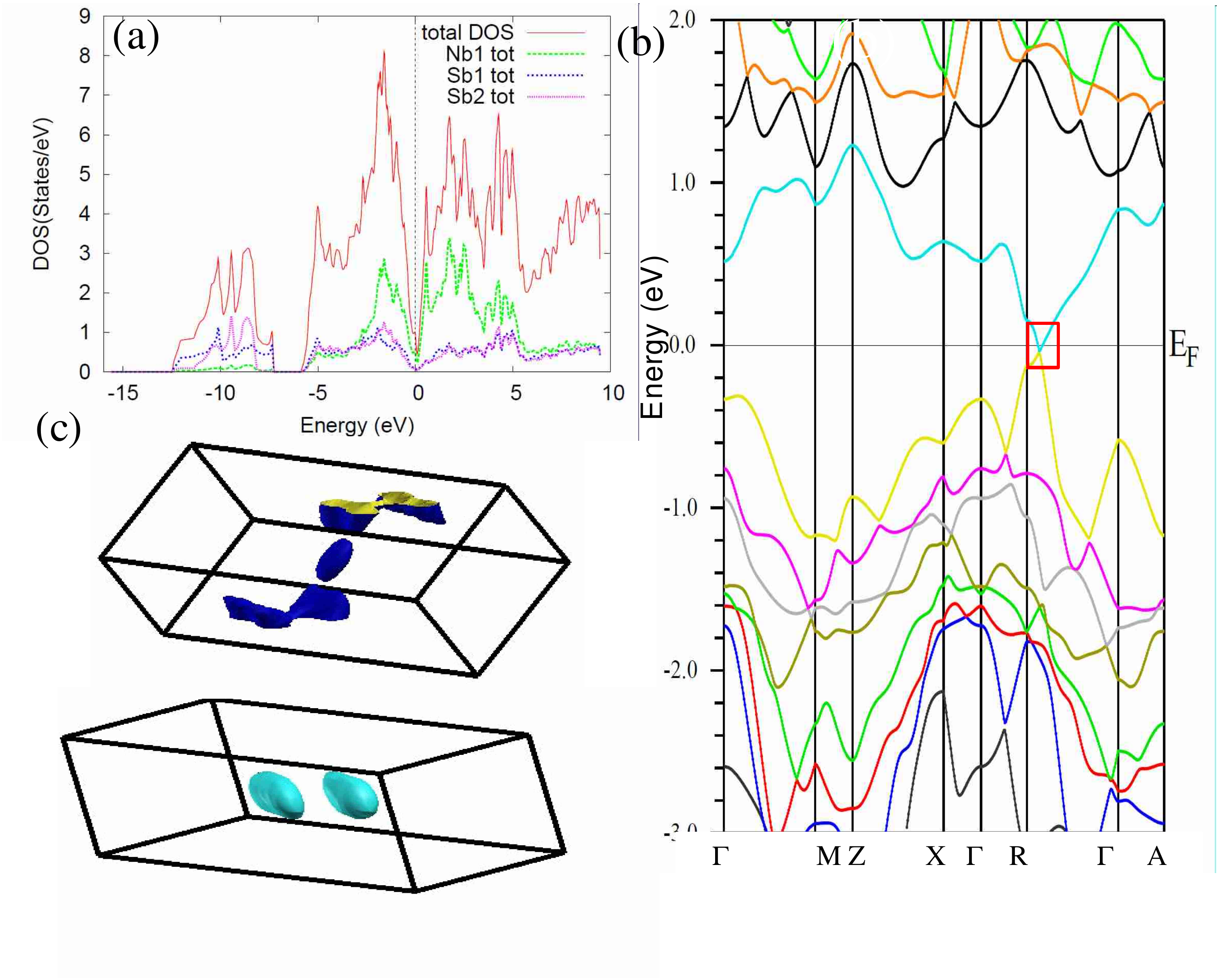}
\caption{(Color online) (a) The total density of states (DOS), as well as the contribution to DOS from different atoms (Nb, Sb1 and Sb2), (b) the band structure and (c) the Fermi surfaces of NbSb$_2$ crystals from first-principle calculation. The spin-orbit coupling was taking into account in process. For clarity the orientation of the hole pockets (the upper panel in (c)) is rotated in comparison to the electron pockets (the lower panel in (c)). The red rectangle in (b) indicates the area where the conduction band touches the valence band at the Fermi level.}
\end{figure}

The Hall resistivity confirms the multiband characteristic of NbSb$_2$. Fig. 3(c) shows the $\rho_{xy}(B)$ in different temperatures with the current parallel to $b$-axis and the magnetic field perpendicular to the current. At 2 K, $\rho_{xy}$ is initially positive below 1 T but changes to negative in higher fields. With increasing temperature, the field where $\rho_{xy}$ changes the sign increases but achieves the maximum (8.5 T) in $\sim$ 150 K. Further increase in the temperature above 150 K induces the decrease of the sign reversal field (Fig. 3(d)). The curvature and sign reversal of the Hall resistivity clearly indicates the coexistence of two types of carriers in NbSb$_2$. This is in agreement with the sign change in Seebeck coefficient in (T,B) (inset in Fig. 3(d)) and with the electronic structure from the first-principles calculation (Fig. 4) which also shows hole and electron pockets. The density of states (DOS) (Fig. 4(a)) reveals that the Fermi level of NbSb$_2$ locates at the valley of the DOS. This is confirmed by the band structure (Fig. 4(b)). The states at the Fermi level are dominated by Nb contribution. The FS of NbSb$_2$ consists of three electron pockets and two hole pockets (Fig. 4(c)). Even though most of the pockets are three-dimensional, there is an electron pocket which is anisotropic at the center of the Brillouin zone and it should dominate the anisotropic magnetoresistance in NbSb$_2$.

Presently there are several possible mechanisms for large unsaturated MR in nomagnetic metals. One is related to the Dirac fermions with linear energy dispersion and possibly originates from the quantum limit \cite{quantummr,qt1,qt2,qt3}. The quantum limit of Dirac fermions gives linear MR \cite{quantummr} which is distinct from the quadratic MR in semiclassical MR \cite{mr3}. However, MR of NbSb$_2$ shows parabolic field-dependence. This implies the quantum limit should not dominate the transport. The low-field-induced increase in $\rho(T)$ in graphite may be caused by excitonic insulator transition of the Dirac fermions \cite{graphite3} or by the FS compensation (equal numbers of electrons and holes) similar to Bi and PtSn$_4$ \cite{graphite2,bi,ptsn4}.

The semiclassical expression for the Hall coefficient including both electron and hole type carriers gives $R_{H}=\frac{1}{e}\frac{(\mu _{h}^{2}n_{h}-\mu _{e}^{2}n_{e})}{(\mu _{e}n_{h}+\mu _{h}n_{e})^{2}}$,
when $\mu _{0}H\rightarrow $ 0, and $R_{H}=\frac{1}{e}\frac{1}{n_{h}-n_{e}}$ when
$\mu _{0}H\rightarrow $ $\infty $, where $e$ is the electron charge, $n_{e(h)}$ and $\mu_{e(h)}$ represent the carrier concentrations and mobilities of the electrons (holes) \cite{mr3}. Once there are two carrier types present, the field dependence of $\rho
_{xy}(H)$ will become nonlinear. However, the sign change of the Hall resistivity in field only occurs when the two kinds of carriers have distinct mobility and density, as observed in Rh \cite{Hall}. The holes with higher mobility and small effective mass dominate and induce the positive Hall resistivity in low field. In order to achieve the sign reversal of Hall resistivity in high field, the number of high mobility carriers (holes) should be much smaller than the number of low-mobility carriers (electrons).

The electronic structure results in Fig. 4(c) reveals distinct volumes of the electron and hole pockets. This is also confirmed by two distinct oscillation frequencies in SdH oscillation (Fig. 3(a)). These observations, taken together with Hall resistivity, imply the inadequacy of FS compensation picture in NbSb$_2$. Furthermore, graphite shows a convex MR behavior in low fields and reentrant metallic behavior in high field, also distinct from our results. The crystalline and electronic structure of NbSb$_2$ also discriminates it from quasi-2D PdCoO$_2$ \cite{PdCoO2-1} and WTe$_2$ \cite{WTe2}.

It was proposed that the magnetic field will induce the breaking of the time reversal invariance and will rearrange the Dirac FS in a Dirac semimetal \cite{semimetal1,semimetal2,semimetal3,semimetal4}. This plus the high mobility of the Dirac carriers could induce very large MR, and is believed to be responsible for the effect observed in Cd$_3$As$_2$ \cite{Cd3As2,Cd3As21,Cd3As22,Cd3As23}. The electronic structure of NbSb$_2$ indeed hosts features similar to a Dirac semimetal. The hole pocket in NbSb$_2$ has very high mobility in comparison to the electron pockets and the effective mass of carriers is very small, as revealed by the quantum oscillation and Hall resistivity. The DOS at the Fermi level is very small (Fig. 4(a)), and the conduction band nearly touches the valence band at only one point at the Fermi level in the band structure whereas the spin-orbit coupling opens a very small gap (Fig. 4(b)). The magnetic field could rearrange the FS so that FS either splits into two disjoint Weyl pockets or two concentric spheres \cite{Cd3As2,semimetal2}. If the Fermi level locates between the two disjoint Weyl pockets, the system will change to a semiconductor, contributing to MR.

In a single band metal with diffusion mechanism and electron-type carriers, Seebeck coefficient in magnetic field is given by the Mott relationship,
\begin{eqnarray}
S(B)&=&\mathcal{A}\left(\frac{\sigma^2}{\sigma^2+\sigma_{xy}^2}\mathcal{D}+\frac{\sigma_{xy}^2}{\sigma^2+\sigma_{xy}^2}\mathcal{D}_H\right)\\
&=&\mathcal{A}\frac{(Ne\mu)^2\mathcal{D}+(Ne\mu^2B)^2\mathcal{D}_H}{(Ne\mu)^2+(Ne\mu^2B)^2};
\end{eqnarray}
where $\mathcal{A}=\frac{\pi^2k_B^2T}{3e}$, $\mathcal{D}=\frac{\partial\ln\sigma}{\partial \zeta}$ and $\mathcal{D}_H=\frac{\partial\ln\sigma_{xy}}{\partial \zeta}$ ($\zeta$ is the chemical potential) \cite{TE1,TE2,TE3}. The Seebeck coefficient is very sensitive to the change of the Fermi surface and was used to probe the Fermi surface reconstruction in cuprate and other superconductors \cite{cuprate1,cuprate2}. In a metal $S \rightarrow 0$ as $T \rightarrow 0$. This is indeed observed for Seebeck coefficient of NbSb$_2$ in zero field (Fig. 3(d) inset). In contrast, the Seebeck coefficient in 9 T field shows a nearly constant value $\sim 2$ $\mu$V/K below 10 K. The nonzero Seebeck coefficient as $T \rightarrow 0$ can be observed in a semiconductor, where Seebeck coefficient $S=\frac{k_B}{e}\left[\frac{5}{2}-\frac{n}{N_c}\right]$ with the conduction band carrier density $n$ from impurity/doping and the intrinsic carrier density $N_c$ \cite{TE1,TE2}. This suggests that a gap may open in NbSb$_2$ in the magnetic field. Whereas our observations are fully consistent with this picture, the response of the FS of NbSb$_2$ to magnetic field deserves further study.

In summary, large magnetoresistance and the magnetic field induced semiconducting gap are observed in NbSb$_2$ single crystal. Transverse $MR$ approaches $1.3\times10^5\%$ in 2 K and 9 T, and $4.3\times 10^6\%$ in 0.4 K and 32 T field without saturation, with electric current parallel to the $b$-axis. The large MR is significantly suppressed when the current is along the $ac$-plane. The Hall resistivity and the first-principle electronic structure calculation indicate the coexistence of a small number of holes with very high mobility and a large number of electrons with small mobility. The observed effects are consistent with the rearrangement of the Fermi pocket with high mobility induced by magnetic field.

\begin{acknowledgments}
We than John Warren for help with SEM measurements. Work at Brookhaven is supported by the U.S. DOE under contract No. DE-AC02-98CH10886 (K. W. and C. P.). Work at the National High Magnetic Field Laboratory is supported by the DOE NNSA DE-FG52-10NA29659 (D. G.), by the NSF Cooperative Agreement No. DMR-0654118 and by the state of Florida.
\end{acknowledgments}


\end{document}